\documentstyle[preprint,prc,eqsecnum,aps,epsf]{revtex}

\tightenlines   

\begin{document}
\draft

\title{Toy Model for Pion Production II: \\ The role of
three-particle singularities}
 
\author{A. Motzke$^a,b$, Ch. Elster$^{a,c}$, and
C. Hanhart$^a$}

\address{
$^a$Institut f\"ur Kernphysik, Forschungszentrum J\"ulich,
D-52428 J\"ulich, Germany \\
$^b$Lehrstuhl f\"ur Theoretische Chemie, Technische
Universit\"at M\"unchen, D-85747 Garching, Germany \\
$^c$ Department of Physics and Astronomy, Ohio University,
Athens, OH 45701, USA}

\date{\today}
 
\maketitle 

\begin{abstract}
The influence of three-particle breakup singularities on $s$--wave meson production in
nucleon-nucleon collisions is studied  within the distorted wave Born approximation.
This study is based on a simple scalar model for the two-nucleon interaction and the
production mechanism. An algorithm for the exact numerical treatment of the inherent three-body
cuts,  together with its straightforward implementation is presented.
It is also shown 
that two often-used approximations to avoid the calculation of the three-body breakup are
not justified.
The possible impact on pion production observables is discussed.
\end{abstract}

\vspace{10mm}



\pagebreak


\section{Introduction}

Interest in studies of pion production in nucleon-nucleon collisions at
energies near the pion production  threshold 
has been revitalized by the appearance of excellent high quality
 data \cite{meyer}.
In addition, experimental data for double polarization observables
for the reactions $pp\to pp\pi^0$ \cite{polpppi0}, $pp\to pn\pi^+$ \cite{polpnpipl},
and $pp \to d\pi^+$ \cite{poldpipl} are available.
A comparison of the experimental information  with the predictions of a modern meson 
exchange model \cite{Han2} shows that the data for the production of
charged pions are well reproduced, whereas there are large discrepancies
between the predictions and the data 
when considering  neutral pion production.
Thus, although the process of pion production has been investigated  over several decades,
 the phenomenology of this fundamental process is still
not fully resolved.

A priori one could assume that the most rigorous approach to pion production
in nucleon--nucleon collisions is a complete description of the nucleon-nucleon-pion
system, e.g. via coupled channel
equations, as developed recently in Ref.~\cite{nnpi}.
However, not only is such a procedure difficult to implement, arguments 
based on effective field theories suggest 
that, at least close to the production threshold, most of the
diagrams generated within a coupled channel approach are suppressed 
\cite{cohen,pwave}. Those arguments indicate that 
the only piece that needs to be  treated non--perturbatively is
the nucleon--nucleon ($NN$) interaction, whereas the transition amplitude  $NN \to NN\pi$ 
can be treated perturbatively. In this spirit, most of the recent calculations
for pion production in $NN$ collisions are based on 
the distorted wave Born approximation (DWBA)
\footnote{Note: Even before using  chiral perturbation
theory, DWBA was used for calculations of  the process $NN \to NN\pi$ \cite{KuR}.
In these calculations the  only justification for this procedure
was based on the slow growth of the $NN$ inelasticities.}.

Above the pion production threshold, where a virtual
pion exchange can produce a real pion,  one encounters a final state containing three 
real particles and therefore has to consider a full three-body breakup amplitude.
This amplitude contains branch cut singularities.
Since the evaluation of those so-called three-body singularities 
is technically rather involved,  they have  been avoided  so far in 
all calculations that apply DWBA to meson production in $NN$ collisions \cite{Han3}.
Instead, the energy--dependence of the pion propagator was manipulated
in such a way
 that no singularity occurred. 
These approximations, however, were never tested quantitatively. Moreover,
since the imaginary part arising from the three-body singularities
scales with the three-body phase space, a strongly energy--dependent
 phase motion
of all those amplitudes with a significant contribution from 
pion exchange should be expected.
One might speculate that the insufficient treatment of the three-body character of the
pion production may 
be a source of the insufficient  description of  the polarization data
for neutral pion production mentioned before.

Recently a simple, scalar `toy' model for pion production was proposed in order to
investigate the validity of several approximations used with respect to the energy
dependence in the pion propagator as well as in the $\pi N$ amplitude \cite{toy}. 
Although the model used in Ref. \cite{toy} is simple with respect to the interactions, 
the pion dynamics is treated exactly.
A significant finding of that work was that
a proper treatment of the dynamics is very important,
even exactly at the pion production threshold.
Therefore,  this model is a natural choice for studying the 
impact of an exact treatment of the
three-body dynamics in the pion production process. 
In this work we specifically  want to address the role of three-body singularities
in reactions of the type $NN \to NN\pi$, and to study their possible effect on observables. 
In addition, we present an algorithm for their evaluation 
that can be implemented quite easily.
For this specific study an exact treatment of the  $NN$ interaction
is not necessary. The inclusion of
a full $NN$ interaction will add to the computational complexity by requiring
careful interpolations of the
half off--shell $NN$ transition amplitudes, but will have no impact on 
the technical peculiarities induced by the three-particle singularities of the
pion propagator. Such a model is thus the ideal starting point for our 
investigation.

This article is structured as follows: In Section~II we briefly summarize the
most important features of the model from Ref.~\cite{toy}; in Section~III the algorithm
for the exact numerical evaluation of the three-body breakup singularity is given; 
in Section~IV we present our numerical results; and we conclude in Section~V.

\section{The `Toy' Model for Pion Production in $NN$ Collisions}

Here the most important features of the scalar model for the production of pions in $NN$
collisions as introduced in Ref.~\cite{toy} are summarized and discussed.
The main features of this model are:

i) The pion is represented by a scalar--isoscalar Klein--Gordon field that couples
via a Yukawa coupling to the nucleons. Only the emission of this particle is considered.

ii) The nucleons are treated as distinguishable.  
As a consequence, pion emission needs to be considered only from one nucleon. 
The symmetric term wherein the pion is emitted from the other nucleon is omitted. 
The simplicity of the model is retained by allowing the pion to
couple to only one of the nucleons. Therefore  pion exchange between
two nucleons is not contained in the model.

iii) Only  pion rescattering by one nucleon is considered. 
This pion rescattering is described by a $\pi N$ seagull vertex 
which is inspired by the
chiral $\pi N$ interaction Lagrangian. 

iv) The nuclear interaction is modeled by 
the exchange of a heavy scalar--isoscalar Klein--Gordon field, named $\sigma$, which
couples to the nucleons
via Yukawa coupling. Since the strength of this coupling 
does not influence the three-body dynamics of the pion production process,
only small couplings are considered for the 
one-$\sigma$ exchange. 

It is typical to treat pion production near threshold using
a non-relativistic expansion in the nucleon momenta.
In the following only  leading terms in this expansion will be considered;
in particular, contributions from anti-nucleons are ignored.

In summary, we consider a scalar model defined by the following
Lagrangian:
\begin{eqnarray}
{\cal L} = \sum_{j=1,2}N_j^\dagger ( i\partial_0 + \frac{\nabla^2}{2M_N})N_j
+ \frac{1}{2}\left[( \partial_\mu \pi )^2 - m_\pi^2 \pi^2 
+ ( \partial_\mu \sigma )^2 - m_\sigma^2 \sigma^2 \right]
\cr
+ \frac{g_\pi}{f_\pi} N_2^\dagger N_2 \pi
+ g_\sigma \sum_{j=1,2}N_j^\dagger N_j \sigma 
+ \frac{C}{f_\pi^2} N_1^\dagger N_1 (\partial_0 \pi)^2 \ .
\end{eqnarray}
Here $M_N$  represents the physical nucleon mass of 939 MeV and
$m_\pi$ is  139 MeV. 
The mass of the $\sigma$ meson is chosen to be 550 MeV.
All diagrams are evaluated at order 
$\frac{g_\pi}{f_\pi} g_\sigma^2 \frac{C}{f_\pi^2}$. 
In the following we do not display these  factors, 
nor other constants that are common to all  amplitudes.
All the loops are finite, so that we do not need to introduce any
regulators.

It is important to point out some of the unrealistic 
features of the model introduced above, even so they will not influence our
study of the three-body dynamics of the pion production process.
We are concerned with near-threshold kinematics, so that 
a scalar particle 
is  produced  in an  $s$--wave, as is the final $NN$ pair. 
Angular momentum conservation requires  that
the initial  $NN$ pair  is also  in an $s$--wave.
In a realistic case,
however, where the pions are pseudoscalar mesons, the production of $s$--wave pions
calls for a $p$--wave in the initial state. Furthermore, the scalar model does not include
a strong short-range repulsion.
Thus the nucleons have stronger overlap within our model, as compared to a 
more realistic treatment. 
However, since the main focus of this work is the three-body
dynamics of the production operator, a study within this model will still give valuable
information. As an aside, the consideration of more realistic interactions would not
introduce new conceptual aspects to the exact treatment of three-body singularities, but
only increase the numerical effort. 

In a DWBA calculation of threshold pion production, any tree level
diagram is modified substantially by the contributions from the initial and
final state interactions. We will therefore concentrate in our model calculations on those
DWBA terms, in which only initial or final state $NN$ interactions are present.
We will thus  ignore the rescattering diagram 
with DWBA contributions in 
{\it both} initial and final $NN$ interactions, 
 since this is a two--loop integration term. Again, 
the initial and final state interactions, 
which occur before or after
the pion rescattering process, are represented by a single $\sigma$ exchange.
All diagrams are evaluated in time-ordered perturbation theory (TOPT). 
The diagrams involving an initial state interaction are depicted in Fig.~\ref{isigraphen}.
In the present work we will only study ladder diagrams, which are given by the
contributions labeled I1 and I2 in that figure. 
There are two additional types of diagrams, namely
the stretched boxes (Fig. \ref{isigraphen}-I3, I4), and 
graphs in which a $\sigma$ is exchanged in between the
emission and rescattering of the virtual pion. 
We ignore those here. Although both groups have  three-body
singularities, they introduce no additional complications and
are thus not relevant for the present study.

\section{Calculation of the Three-Body Breakup}

As mentioned at the end of the former section we will study only ladder diagrams.
 The corresponding diagrams are shown 
in Fig.~2 together with the relevant momenta.
In time-ordered perturbation theory the corresponding amplitudes are proportional to
\begin{eqnarray}
\label{misi1} 
A^{F} & = & \int d^3 p'' \; V_\sigma^F
\; G_{NN}^F(G_{\pi NN}^{F,r}-G_{\pi \pi NN}^{F,b}) \ , \\
A^{I} & = & \int d^3 p'' \;
\; (G_{\pi NN}^{I,r}-G_{\pi \pi NN}^{I,b}) G_{NN}^IV_\sigma^I
 =:A^{I,r}+A^{I,b} \ .
\label{misi2} 
\end{eqnarray}
The superscript $r$ ($b$) denotes the diagrams 
for pion rescattering (backscattering), namely those
with a pion moving
forward (backward) in time, as shown in Fig.~1 I1 (I2).
The individual pieces can be directly deduced from the diagram;
e.g. the two propagators in Eq.~(\ref{misi2}) are given by
\begin{equation}
G_{\pi NN}^{I,r} =  \left( \frac{1}{E_{tot}- E_+' - E'' -\omega_q +i\epsilon} \right) \ ,
\label{gpinndef}
\end{equation}
and
\begin{equation}
G_{\pi \pi NN}^{I,b} =  \left( \frac{1}{E_{tot}- E_-' - E'' -\omega_q-\omega_{q'} +i\epsilon} \right).
\label{gpinnbackdef}
\end{equation}
 The relative
minus sign between the two propagators in Eqs. (\ref{misi1}) 
and (\ref{misi2})
 stems from the 
particular form of the $\pi N \to \pi N$ transition operator, namely
the pion energies that appear explicitly.
The evaluation of the expressions for
$A^F$ as well as $A^{I,b}$ is straightforward since they only contain two
nucleon singularities, as long as total energies $E_{tot}$
 below the two-pion threshold are considered (c.f. Eq. (\ref{gpinnbackdef})).
Thus in the following we will concentrate on the evaluation of the amplitude $A^{I,r}$. 
When discussing the results, however, the complete amplitudes $A^F$ and as $A^I$ will
be considered.

For completeness we will also give explicitly the
expressions for the other parts of $A^{I,r}$, namely
the two-nucleon propagator
\begin{equation}
G_{NN}^I=\frac{1}{E_{tot}- 2E''+i\epsilon} \; , 
\end{equation}
and the $\sigma$-exchange potential
\begin{equation}
 V_\sigma^I =
 \frac{1}{\omega_\sigma(E_{tot} - E -  E'' - \omega_{\sigma})} \; .
\label{Vsigi}
\end{equation}
In a model based on a realistic $NN$ interaction, the potential 
$V_\sigma^I$ will have to be replaced by a $NN$ t-matrix.

The integration variable  $p''$ in Eq.~(\ref{misi2}) stands for 
 the relative momentum of the nucleons in the intermediate state.
The quantity $E_\pm' = \frac{1}{2M_N}(\vec{p} \,' \pm \frac {\vec{q} \,'} {2})^2 $ is
the energy of the right ($+$) and the left ($-$) nucleon in the final state,
 $\omega_{q'} = \sqrt{m_\pi^2+\vec q{}\,'^2}$ is the energy of the produced pion,
and the total energy is given by  $E_{tot} = 2E = E_+' + E_-' + \omega_{q'}$.
In addition, 
$\omega_q = \sqrt{m_\pi^2+\vec q{}\,^2}$ and $\omega_\sigma
 = \sqrt{m_\sigma^2+\vec k{}^2}$ denote the on-shell energies
of the $\pi$ and $\sigma$ meson, and 
 $E''=\frac{{\vec p}\, ''^2}{2M_N}$ is the energy of a nucleon in an
intermediate state. The  momenta $\vec{q}$ and $\vec{k}$ are defined through
$\vec{q} = \vec{p}\,' + \frac{\vec{q}\,'}{2} - \vec{p}\,''$
and $\vec{k} = \vec{p} - \vec{p}\,''$
. The momentum of the initial nucleon is labeled
$\vec{p}$ (c.f. Fig. \ref{koord2}).

The amplitude given by Eq. (\ref{misi2}) exhibits a distinct singularity structure; it
contains both, a two- and a three-body cut. Those cuts occur whenever it is energetically
allowed for two or three particles in the intermediate state to go on-shell simultaneously.
It should be noted that the two and three-body singularities are kinematically well separated,
as shown in detail in Appendix A;
whenever the intermediate two-nucleon state is on-shell, the $\pi NN$ propagator
is strictly negative. 
The numerical treatment of two-body singularities, e.g. by subtraction methods, is a
well-established procedure \cite{hafteltabakin}. 
Calculations of the three-body breakup have long history in neutron-deuteron scattering
(e.g. \cite{gloecklebook,witala,takejami}). 
In the context of $NN$ potentials that incorporate 
pion production explicitly, the breakup has been treated either by complex contour
deformation \cite{cenn}, spline methods \cite{leemat}, 
or subtraction methods \cite{faassen,Schwamb}. For 
the explicit calculation of pion production observables the last seem preferable, since
all calculations are carried out along the real axis. For our purpose, namely 
 the framework of DWBA, the method introduced by M. Schwamb
\cite{Schwamb,Schwambthesis} is most suited.
Thus, this algorithm is presented here together with modifications necessary
for the evaluation of the loop given in Fig.~2.

In order to proceed,  the $\pi NN$ propagator introduced in Eq. (\ref{gpinndef}) is
rewritten as
\begin{equation}
G_{\pi NN}^{I,r}= \left( \frac{E_{tot}- E_+'-E''+\omega_q}{2P'p''}\right)
 \frac{1}{x''-x_0+i\epsilon} \ .
\label{propumf}
\end{equation}
Here 
$\vec{P}\,' = \vec{p}\,' + \frac{\vec{q}\,'}{2}$ \ and
\begin{equation}
\omega_{q} = \sqrt{m_\pi ^2+{P\,'}^2+{p\,''}^2-2P\,'p\,''x\,''} \ ,
\end{equation}
where $x\,''=(\vec{P}\,'\cdot \vec{p}\,'')/(P \,'p \,'')$.
The quantity $x_0$ is defined as
\begin{equation}
x_0 \equiv x_0(P\,',p\,'' ) =
\frac {m_\pi^2+{P\,'}^2+{p\,''}^2-(E_{tot}-E_+'-E'')^2}{2P\,'p\,''} \ .
\label{xxx0}
\end{equation}

The integration over the angle variable $x\,''$ contains a singularity for all $p\,''$
that leads to $|x_0| \le 1$. Thus,
\begin{eqnarray}
x_0=\pm1 &\Leftrightarrow&
m_\pi^2+(P\,'\mp p\,'')^2 -
\left( E_{tot}-\frac {{P\,'}^2} {2M_N}-\frac {{p\,''}^2} {2M_N} \right)^2=0
\label{polynom1}
\end{eqnarray}
represent two conditions for the boundaries of the area containing three-particle
singularities. These are two polynomial equations of degree four for
$p\,''=p\,''(P\,')$. Only three of the eight solutions are physically relevant,
as can be easily checked by taking the threshold limit for the solutions. They 
enclose the area of three-particle singularities, as illustrated in
Fig.~\ref{mond}, for a total energy of
$E_{tot}=210$~MeV. In what follows we will denote the boundaries of 
this area by $p_a$ and $p_b$, respectively, as indicated in the figure.

Introducing the auxiliary function
\begin{equation}
F(x\,'') := (E_{tot}- E_+'-E''+\omega_q)
\int_0^{2\pi}d\varphi \,'' \; V_\sigma^I
\label{F}
\end{equation}
allows to rewrite the amplitude $A^{I,r}$ from Eq.~(\ref{misi2}) as
\begin{eqnarray}
A^{I,r}= A_{3b}^I + A_{no}^I+ A_{2b}^I \ , 
\label{Fint}
\end{eqnarray}
where
\begin{eqnarray}
A_\alpha^I&=&\frac{M_N}{2P'} \int_{R_\alpha} dp\,'' \frac {p\,''} {p^2-{p\,''}^2+i\epsilon}
\int_{-1}^1 dx\,'' \frac {F(x\,'')} {x\,''-x_0+i\epsilon} \ . 
\end{eqnarray}
The regions of integration are given by
\begin{eqnarray}
R_{3b} = [p_a,p_b] \ ; \qquad R_{no} = [0,p_a] \ \cup \ [p_b,p_i] \ ;
\ \qquad R_{2b} = [p_i,\infty] \ .  
\label{intdef}
\end{eqnarray}
Here $p_i$ denotes an in-principle arbitrary momentum placed in the interval
between the region of three-body singularities and the two-nucleon 
singularity 
(c.f. Fig. \ref{mond}). For the numerical
evaluation we chose $p_i$ such that it cuts this interval in half.

The term $A_{3b}^I$ is thus the piece of the amplitude that contains the three-body cut.
The second term $A_{2b}^I$ in Eq.~(\ref{Fint}) stands for  the  remaining
integration from $p_i$ to infinity and contains only the 
two-nucleon singularity. Since this calculation involves only a standard, single
subtraction it will not be discussed in detail.
The third term, $A_{no}^I$, contains the integration outside the 
region of three-body singularities for momenta smaller than $p_i$
and will be discussed in detail at the end of this section.

We begin our discussion with the term $A_{3b}^I$.
For each value of $x_0$ we subtract and add the value of $A_{3b}^I$ at the singularity and get
\begin{eqnarray}
\nonumber
A_{3b}^I&=&\frac{M_N}{2P'}\int_{p_a}^{p_b} dp\,'' \frac {p\,''} {p^2-{p\,''}^2}
\\
&& \ \times \left[ \int_{-1}^1 dx\,'' \frac {F(x\,'')-F(x_0)} {x\,''-x_0} +
 F(x_0) \left( \ln \left| \frac{1-x_0}{1+x_0} \right| - i\pi \right) \right] \ .
\label{subisi}
\end{eqnarray}
The complication specific to three-body singularities is the appearance of 
removable logarithmic singularities: the arguments
of the logarithm vanish at the  integration limits
$p_a$ and $p_b$, where $|x_0|=1$. 

The last piece of Eq. (\ref{subisi}) is the imaginary part mentioned in the
introduction. It leads to an additional imaginary contribution, which is necessarily 
energy--dependent.

Let us concentrate now on the logarithm in the  last term of Eq.~(\ref{subisi}). Here we employ a
transformation first introduced by Schwamb \cite{Schwambthesis} that allows to carry out the
integrals numerically in a quite straightforward fashion. In order to apply the transformation
the range of  integration over $p''$ is split into two pieces, so that
there is only one singularity in each of the two intervals.
The intermediate momentum is chosen to be  $p_{ab}:=\frac{1}{2}(p_a+p_b)$,
in accordance with Ref. \cite{Schwambthesis}.
Using the  abbreviations $\alpha(p\,'')$ and
$\beta(p\,'')$, we are now faced with integrals of the type 
\begin{equation}
{\cal{I}} := \int_{p_{ab}}^{p_b} dp\,'' \alpha(p\,'') \, \ln [\beta(p\,'')]
\label{lnint}
\end{equation}
where $\beta(p_b)=0$. The  substitution 
\begin{equation}
y^2 := p_b-p\,'' \ge 0
\label{schwambtrafo}
\end{equation}
allows to transform the integral of Eq. (\ref{lnint}) into
\begin{equation}
{\cal{I}} = 2 \int_0^{\sqrt{\frac{p_b-p_a}{2}}} dy
\, y \, \alpha(p_b-y^2) \, \ln [\beta(p_b-y^2)] \ .
\label{lnint2}
\end{equation}
Due to  the minus sign in Eq.~(\ref{schwambtrafo}) the lower
integration limit $0$ is now the `critical' point of the integrand.
The additional factor $y$ in Eq.~(\ref{lnint2}) is responsible for the `disappearance' 
of the logarithmic singularity; the integral is now regular, since
$\lim_{y\to 0} y \ln (y) = 0$. The integration interval
 $[p_a,p_{ab}]$ is treated in a similar fashion, as described in Appendix B. 

In principle, the application of a subtraction
in the intervals $[0,p_a]$ and $[p_b,p_i]$ is not necessary, since 
singularities occur at the boundaries only. However, numerical tests showed that
applying a subtraction in these intervals also allows a smaller number of grid
points in obtaining a stable and convergent result. 
In summary, with the above algebraic transformations, the second term in Eq.~(\ref{Fint})
is expressed as
\begin{eqnarray}
\nonumber
A_{no}^I&=&\frac{M_N}{2P'}\int_{R_{no}} dp\,'' \frac {p\,''} {p^2-{p\,''}^2} \\
&& \ \ \times \ \left[ \int_{-1}^1 dx\,'' \frac {F(x\,'')-F(\mbox{sign}\,x_0)} {x\,''-x_0} +
F(\mbox{sign}\,x_0) \ln \left| \frac{1-x_0}{1+x_0} \right| \right] \; ,
\label{subisi2}
\end{eqnarray}
where the integration interval $R_{no}$ was defined in Eq. (\ref{intdef}).

Note that, through the splitting into intervals of the $p ''$ integration 
the first term in Eqs. (\ref{subisi}) and (\ref{subisi2}) also appears as
integrated in intervals.
In principle there are no numerical difficulties with this integral, however one needs to make
sure that there is a sufficient number of  mesh points  within the region
of the three-body singularities in order to obtain accurate results.
That is why we propose to use the splitting of the intervals  here as well.
A complete presentation of the three-body cut algorithm is given in 
Appendix B.

\section{Effect of the Three-Body Breakup on the Production Cross Section}

Since the ingredients of our model are of scalar nature, the only observables to be considered
here are cross sections. The total cross section $\sigma_{tot}$ for pion production
is calculated from the  coherent sum of the production amplitudes $A^F$ and $A^I$ defined
in Eqs. (\ref{misi1}) and (\ref{misi2}), respectively. Here we will use the notation
\begin{equation}
\sigma_{tot} = \int d\rho |A^I+A^F|^2 \quad ; \qquad
\sigma_{tot}^F = \int d\rho |A^F|^2 \quad ; \qquad
\sigma_{tot}^I = \int d\rho |A^I|^2 \quad ,
\label{eq:4.1}
\end{equation}
where $d\rho$ denotes the phase space factor.
As a measure of the total energy we introduce the excess energy $Q = E_{tot}-m_\pi$.
 In the present work
we concentrate on excess energies between the one-pion and two-pion threshold. Thus, if the
excess energy is given in units of $m_\pi$, we restrict ourselves to 
 values of $Q/m_\pi$ between 0 and 1. 
In the upper panel of
 Fig.~\ref{res1} we display the result of the full calculation, $\sigma_{tot}$ (solid line),  
with cross sections $\sigma_{tot}^I$
(dashed line) and $\sigma_{tot}^F$ (dotted line) obtained from the initial
and final state interactions individually.

 In order to  emphasize better the contributions of the FSI and
ISI separately, we divide $\sigma^I_{tot}$ and $\sigma^F_{tot}$ by $\sigma_{tot}$
 and  display those ratios in the lower panel of Fig.~\ref{res1}.   
The figure shows that $\sigma^I_{tot}$ gives the largest contribution
 over the entire energy interval under
consideration, while $\sigma^F_{tot}$ is always close to $\sigma_{tot}$. 

Naturally these features are specific to the model. More interesting is to compare the
exact calculation with approximations for the pion propagator 
 to avoid the appearance of the three-body singularities, since a main
goal of this work is to investigate the quality of those approximations. Here we consider
two of these approximations. In both, the $\pi NN$  propagator 
Eq.~(\ref{gpinndef}) is substituted by a propagator containing no three-body singularity.
In case of the \emph{static approximation} (used, e.g., in Ref. \cite{static})
the substitution
\begin{equation}
\frac{1}{E_{tot}- E_+' - E'' -\omega_q +i\epsilon} \: \longrightarrow 
 -\frac{1}{\omega_q}
\label{static}
\end{equation}
is made; i.e., the  $\pi$-exchange is made instantaneous.
In case of the `\emph{threshold approximation}' (used in Refs. \cite{Han1,thresha}),
 the $\pi NN$ propagator
is given by
\begin{equation} 
\frac{1}{E_{tot}- E_+' - E'' -\omega_q +i\epsilon} \: \longrightarrow
    \frac{1}{m_\pi-E_+'- E''-\omega_q} \ .
\label{thresholdap}
\end{equation}
By definition, exactly at threshold this `threshold approximation'
agrees with the exact propagator, since the total energy $E_{tot}$
is then equal to $m_\pi$. In the literature other approximations
can be found. For example, in Ref. \cite{cohen} a propagator similar to
that of Eq. (\ref{thresholdap}) is used, but with $E_+' \equiv m_\pi/2$ and
$E'' \equiv 0$. Recognizing that our list is not complete, we nevertheless consider 
the `static approximation' and the 'threshold approximation' as
representative and  concentrate on these.
 
In Fig.~\ref{res3} the ratios of the calculations based on the two approximations 
and the exactly calculated
cross sections  are displayed. 
In panel (c) of Fig.~\ref{res3} the results of the
full calculations (i.e. containing FSI and ISI) are shown. The dashed line gives the
result for the `static approximation', while the dotted line represents the `threshold
approximation'. The figure indicates that, although the `threshold approximation' is by 
definition
exact at threshold, its energy behavior does not correspond to the exact calculation at
higher energies. Especially when approaching the two-pion threshold, this
approximation underestimates the full result considerably. 
The `static approximation' overestimates the full
result in the entire energy regime between the one-pion and two-pion thresholds.
This is especially pronounced near the one-pion threshold.
In order to investigate the effects of the approximations in more detail we display  
their effects calculated with FSI and ISI only in panels (a) and (b) of Fig.~\ref{res3}.
Comparing the panels (a) and (b) with the complete calculation (c) shows that the 
 last is the result of interference effects: due to their singularity structure
$A^I$ and $A^F$ have different phases with different energy--dependence that are 
taken into account
in different ways by the two approximations.

In order to study this point in more detail
let us now consider  the phase motion introduced by the exact inclusion
of the three-body singularities. Obviously, in order to obtain a quantitative
understanding, 
we need to compare the full result to  one in which  the breakup does not occur.
For this comparison we choose the `static approximation', defined in Eq. (\ref{static}). 
In Fig.~\ref{res4} we
compare the phase $\phi [A]$ of the amplitudes, defined through
\begin{equation}
\tan (\phi [A] ) = \frac{\mbox{Im}(A)}{\mbox{Re}(A)} \; ,
\label{eq:4.4}
\end{equation}
for a fixed total excess energy $Q=\frac{1}{2}m_\pi$ as a function of the  kinetic
energy of the relative motion
 of the outgoing two nucleon system. The solid line shows the phase of
the full result, while the dashed line corresponds to the `static approximation'.
panel (a) contains the phase  $\phi [A^F]$, panel (b) the phase $\phi [A^I]$, and
 the phase
of the full calculation, $\phi [A^F+A^I]$ , is shown in  panel (c).
As could have  been expected, the phase of $A^I$ is influenced most by the approximation,
and there is a significant effect on the  phase of the total amplitude. We would like to point
out that for each value of $Q$ that is smaller than $m_\pi$, the effect is qualitatively
similar to the one shown in Fig.~\ref{res4}. Naturally,
this effect is irrelevant as long as only one partial wave contributes. However,
as soon as differential observables are analyzed, different partial waves
start to interfere. It will be important to see how such a change in the phases
of some amplitudes influences the description of the polarization data
of the reaction $NN \to NN\pi$.

\section{Summary and Conclusions}

The effect of  three-body dynamics in the pion production process
within the framework of the DWBA approximation has been investigated 
using a scalar `toy' model. This model is simple with
respect to the interactions employed: the $NN$ interaction is represented by the exchange of a
scalar $\sigma$-meson; the pion is also treated as a scalar, and finally the nucleons
(represented by scalar fields) are assumed to be distinguishable. However, the underlying
dynamics is treated exactly, especially the three-body breakup $NN \rightarrow NN \pi$. We
presented an algorithm for evaluating the three-body breakup singularities exactly. This
algorithm can be applied to realistic $NN$ interactions in a straightforward way. 

Within our model we compared the results for the total production cross section with two
often-used approximations: the `static approximation', which makes the pion exchange
instantaneous, and the `threshold approximation', which fixes the total energy 
at the threshold value. It turns out that both approximations are different in character, but
equally unsuited to describe the total pion production cross section between the one- and
two-pion thresholds, at least within this model study. In case of the static 
approximation the cross section is overestimated
over the entire energy regime, but  especially close to threshold. Although the threshold
approximation is exact at the threshold, it overestimates the cross section slightly close to
threshold and underestimates it considerably near the two-pion threshold. Both
approximations produce the wrong energy--dependence in the energy region under consideration.
It remains to be seen how much of these differences survive in a realistic calculation, 
where, e.g., the final state interaction plays an a lot more prominent
role and small momenta are suppressed due to chiral symmetry.

Finally, we compared the phases of the amplitudes of our exact calculation with the phases
given by the `static approximation' and found sizable differences.
 Although not presented in this
work, a quite similar statement can be made with respect to the `threshold approximation'.
Naturally, the phases play an important role when different partial waves interfere,
as occurs for  differential observables. It can then  be expected that in the case of realistic
interactions the description of some observables will depend crucially on the way that the
three-body dynamics of the intermediate states is treated.


\vfill

\acknowledgements 
We thank J. Durso for careful reading of the manuscript.
This work was performed in part under the
auspices of the U.~S.  Department of Energy under contract
No. DE-FG02-93ER40756 with Ohio University.

\newpage

\appendix

\section{Location of the two and three-body singularities}

We demonstrate here that the two- and three-body
singularities from Eq.~(\ref{misi2}) are well separated.
The two particle singularity occurs at
\begin{eqnarray}
p\,'' = p = \sqrt{M_N E_{tot}},
\label{ucut1}
\end{eqnarray}
i.e., in  case of on--shell scattering.
An estimate for the denominator of the three-body propagator 
clarifies the position of its singularities, namely
\begin{eqnarray}
\nonumber
\forall \quad \tilde{p} \ge p: &&
E_{tot}- E'_+ -E''(p''=\tilde{p})-\omega_q(p''=\tilde{p}) \\
&\le& \frac {E_{tot}} {2} - E'_+ - \omega_q(p''=\tilde{p}) \;\, < \;\, 0 \ ,
\label{ucut2}
\end{eqnarray}
since $E''(p''=\tilde{p}) \ge E=\frac{E_{tot}}{2}$, and $E_{tot}<2m_\pi$,
$\omega_q\ge m_\pi$, and $E'_+ = \frac{{P'}^2}{2M_N} \ge 0$. Consequently the
$\pi NN$--propagator is regular for all $\tilde{p} \ge p$. In particular, the position
of the two-nucleon pole occurs \emph{above} each three-particle singularity.
This leads to the natural decomposition of the radial integration interval
$\int_0^\infty = \int_0^{p_i} + \int_{p_i}^\infty$
 (c.f. Eq. (\ref{Fint})), where
$[0,p_i]$ contains all three-body singularities and $[p_i,\infty]$ contains the
two-body singularity.
\section{Analytic and Numeric Treatment of the Logarithmic Singularities}
Here we give a brief but complete presentation of the three-body cut algorithm.
With
\begin{eqnarray}
\xi_0:= \left\{ \begin{array}{ll}
x_0 & \textrm{if $|x_0| \le 1$} \\
\mbox{sign} \, x_0 & \textrm{if $|x_0| > 1$}
\end{array} \right.
\label{xi0}
\end{eqnarray}
(i.e., $\xi_0=x_0$ in $R_{3b} = [p_a,p_b]$ and $\xi_0=\mbox{sign} \, x_0$
in $R_{no} = [0,p_a] \ \cup \ [p_b,p_i]$,) the contribution
$A_{3b}^I + A_{no}^I$ (c.f. Eqs. (\ref{Fint}), (\ref{subisi}), and (\ref{subisi2}))
to the production amplitude reads
\begin{eqnarray}
A_{3b}^I + A_{no}^I &=&
A^{I,\, \ln} + A^{I,\, \Delta} + i \, \mbox{Im}(A_{3b}^I) \ ,
\label{mdisi2}
\end{eqnarray}
where
\begin{eqnarray}
\mbox{Im}(A_{3b}^I)
&=& - \frac{\pi M_N}{2P'}
\int_{p_a}^{p_b} dp\,'' \frac {p\,''F(x_0)} {p^2-{p\,''}^2} \ ,
\label{imm}
\end{eqnarray}
\begin{eqnarray}
A^{I,\, \ln} &:=& \frac{M_N}{2P'}
\left( \int_{0}^{p_a} + \int_{p_a}^{p_{ab}} +
       \int_{p_{ab}}^{p_b} + \int_{p_b}^{p_i} \right)
dp\,'' \frac {p\,''F(\xi_0)} {p^2-{p\,''}^2}
\ln \left| \frac{1-x_0}{1+x_0} \right| \ ,
\label{mln}
\end{eqnarray}
and
\begin{eqnarray}
A^{I,\, \Delta} &:=& \frac{M_N}{2P'}
\left( \int_{0}^{p_a} + \int_{p_a}^{p_b} + \int_{p_b}^{p_i} \right)
dp\,'' \frac {p\,''} {p^2-{p\,''}^2} \int_{-1}^1 dx''
\frac {F(x'')-F(\xi_0)} {x''-x_0}
\ .
\label{mrest}
\end{eqnarray}
$A^{I,\, \ln}$ denotes the real contribution to
$A_{3b}^I + A_{no}^I$ containing the logarithm
originating in the principal value of the analytically solvable integral,
and $A^{I,\, \Delta}$ is the other real contribution originating in the
application of the subtraction method. 
The expression $A^{I,\, \ln} =: \frac{M_N}{2P'} \sum_{j=1}^4 {\cal{I}}_j$ from
Eq. (\ref{mln})
is to be understood as if the corresponding Schwamb's substitution would have been
already performed in each of the four integrals (c.f. Table \ref{table1}).

\pagebreak



\begin{table} \caption{\label{table1}
Schwamb's substitutions in the four parts of the interval $[0,p_i]$.}

\begin{tabular}{ccc}
interval & transformation & contribution ${\cal{I}}_i$ \\  \hline
$[0,p_a]$ & $y^2 := p_a-p\,'' \ge 0$ &
$-2 \int_{\sqrt{p_a}}^0 dy
\, y \, \alpha(p_a-y^2) \, \ln [\beta(p_a-y^2)]$ \\
$[p_a,p_{ab}]$ & $y^2 := p\,''-p_a \ge 0$ &
$2 \int_0^{\sqrt{p_{ab}-p_a}} dy
\, y \, \alpha(p_a+y^2) \, \ln [\beta(p_a+y^2)]$ \\
$[p_{ab},p_b]$ & $y^2 := p_b-p\,'' \ge 0$ &
$-2 \int_{\sqrt{p_{ab}-p_a}}^0 dy
\, y \, \alpha(p_b-y^2) \, \ln [\beta(p_b-y^2)]$ \\
$[p_b,p_i]$ & $y^2 := p\,''-p_b \ge 0$ &
$2 \int_0^{\sqrt{p_i-p_b}} dy
\, y \, \alpha(p_b+y^2) \, \ln [\beta(p_b+y^2)]$ \\
\end{tabular}
\end{table}

\pagebreak

\noindent
\begin{figure}
\caption{The six TOPT diagrams of the initial state interaction. A pion field is
represented by a dashed line, the nucleon by a single solid line.
The two possible time orderings of the $\sigma$--exchange (solid double line) lead to
the identical contributions grouped in I1 and I2, for each of two ladder diagrams.
I3 and I4 are stretched boxes.
\label{isigraphen}}
\end{figure}

\noindent
\begin{figure}
\caption{Feynman diagram and choice of coordinates  
for a) final-- and b) initial--state interaction. 
 All momenta indicated are three--momenta.
\label{koord2}}
\end{figure}

\noindent
\begin{figure}
\caption{ The region of three-body singularities
for $E_{tot}$=210~MeV and the choice
of the integration intervals for the three-body cut algorithm.
Inside the region enclosed by the solid line we have $|x_0|<1$
and on
the boundary $|x_0|=1$.
Additionally, the two-particle singularity at $p\simeq 444$~MeV is marked.
The symbol $+$ ($-$) means that the corresponding boundary curve is solution
of $x_0=+1$ ($x_0=-1$) (c.f. Eq.~(\ref{polynom1})). 
\label{mond}}
\end{figure}

\noindent
\begin{figure}
\caption{The total production cross section as function of the excess energy, which is given
in units of $m_\pi$. In the upper panel (a) the full calculation is represented
by the solid line, the calculation with the FSI (ISI) only is shown as dotted (dashed) line. 
The lower panel (b) shows the ratios of the
calculations containing the FSI (dashed) and ISI (dotted) only to the full calculation
(solid line in panel (a)).\\
\label{res1}}
\end{figure}

\noindent
\begin{figure}
\caption{ The ratios of the total cross section of the static
approximation (dashed curve) and the `threshold approximation'
(dotted curve) to the total cross section of the corresponding exact calculation.
The calculations of  panel (a) contain only the FSI, the calculations of panel (b) only
the ISI. In panel (c) the ratios of the full calculations are shown.
\label{res3}}
\end{figure}

\noindent
\begin{figure}
\caption{ The phase of the production amplitude in degrees as a function
of the kinetic energy of the relative motion of the $NN$--system ($\varepsilon$ in
units of $m_\pi$) at the excess energy
$Q=\varepsilon_{max}=\frac{1}{2}\,m_\pi \simeq 70\mbox{MeV}$. The solid lines
represent the exact calculation, the dashed lines 
the `static approximation'. 
In panel (a) calculations with the FSI only are shown, in panel (b) those with the
ISI only. The full calculations, i.e. the superposition of (a) and (b), are shown in
panel (c).  Qualitatively, for each $Q$
with $0\leq Q< m_\pi$ the illustrated dependence is the same.
\label{res4}}
\end{figure}

\newpage

\input{psfig}

{\bf Fig.~1}

\vspace{10mm}

\centerline{\psfig{file=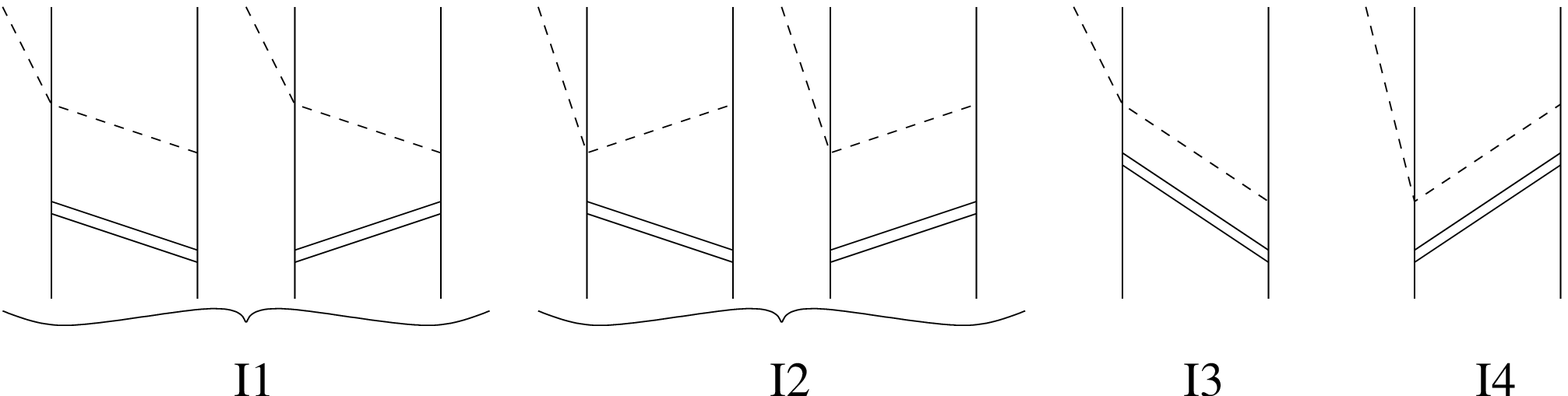,width=14.cm}}

\vspace{20mm}
 
{\bf Fig.~2}

\vspace{10mm}

\centerline{\psfig{file=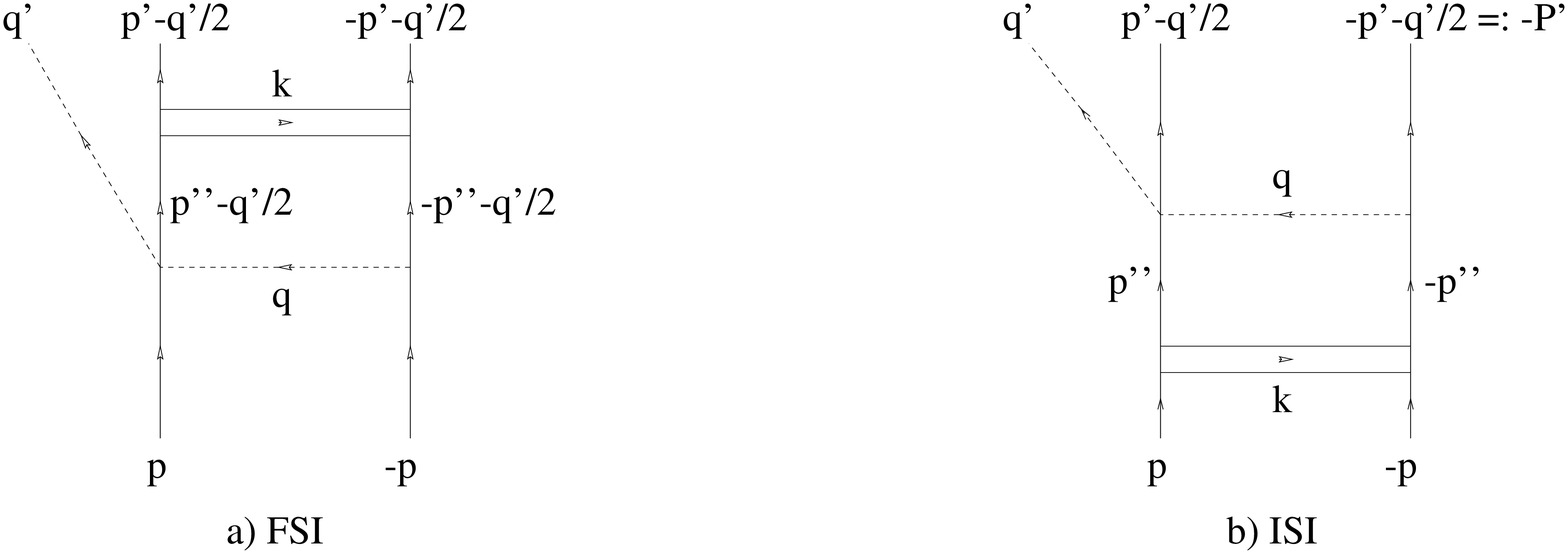,width=15.5cm,height=6cm}}

\newpage

{\bf Fig.~3}

\centerline{\psfig{file=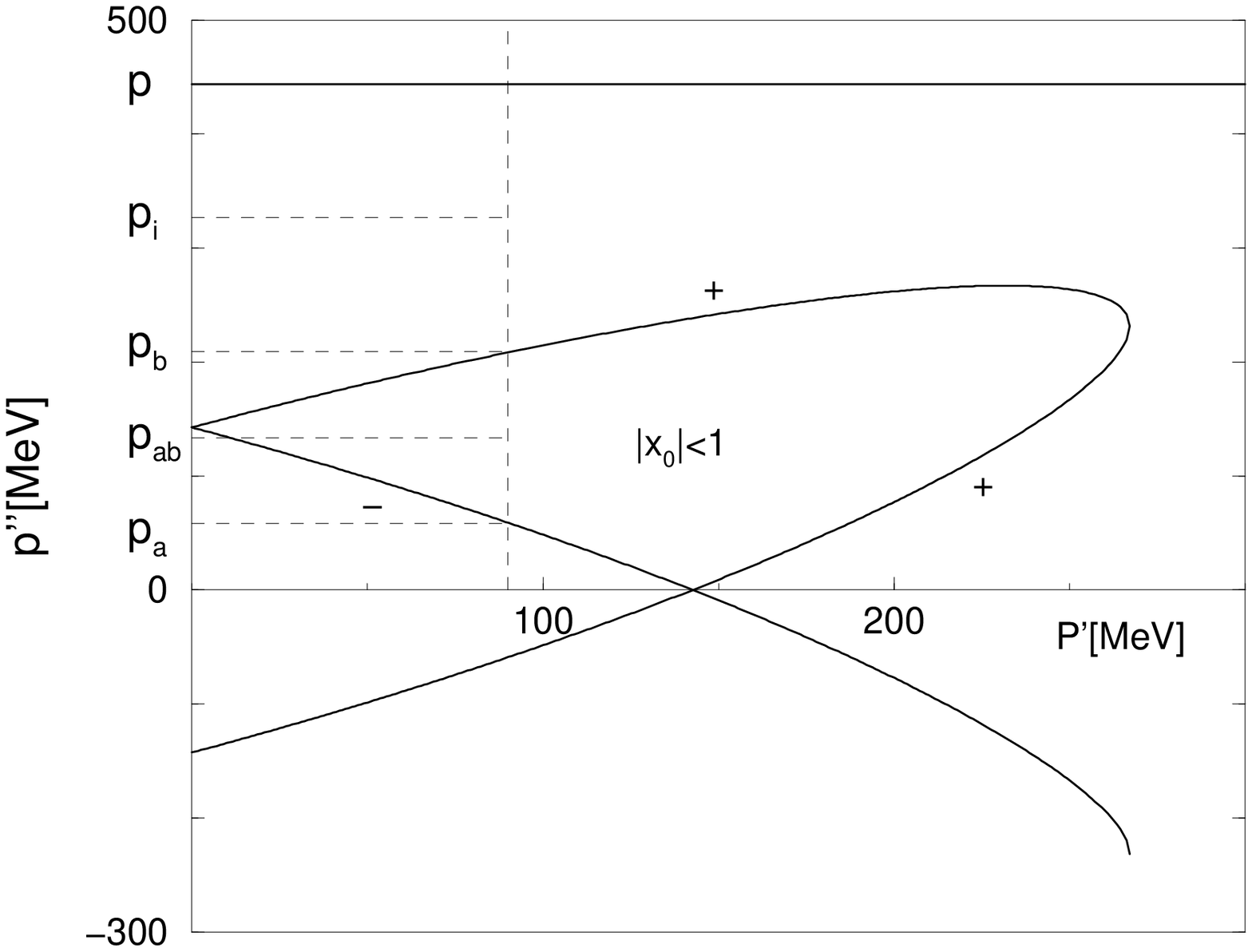,height=7cm,width=7cm}}

\vspace{20mm}

{\bf Fig.~4}

\centerline{\psfig{file=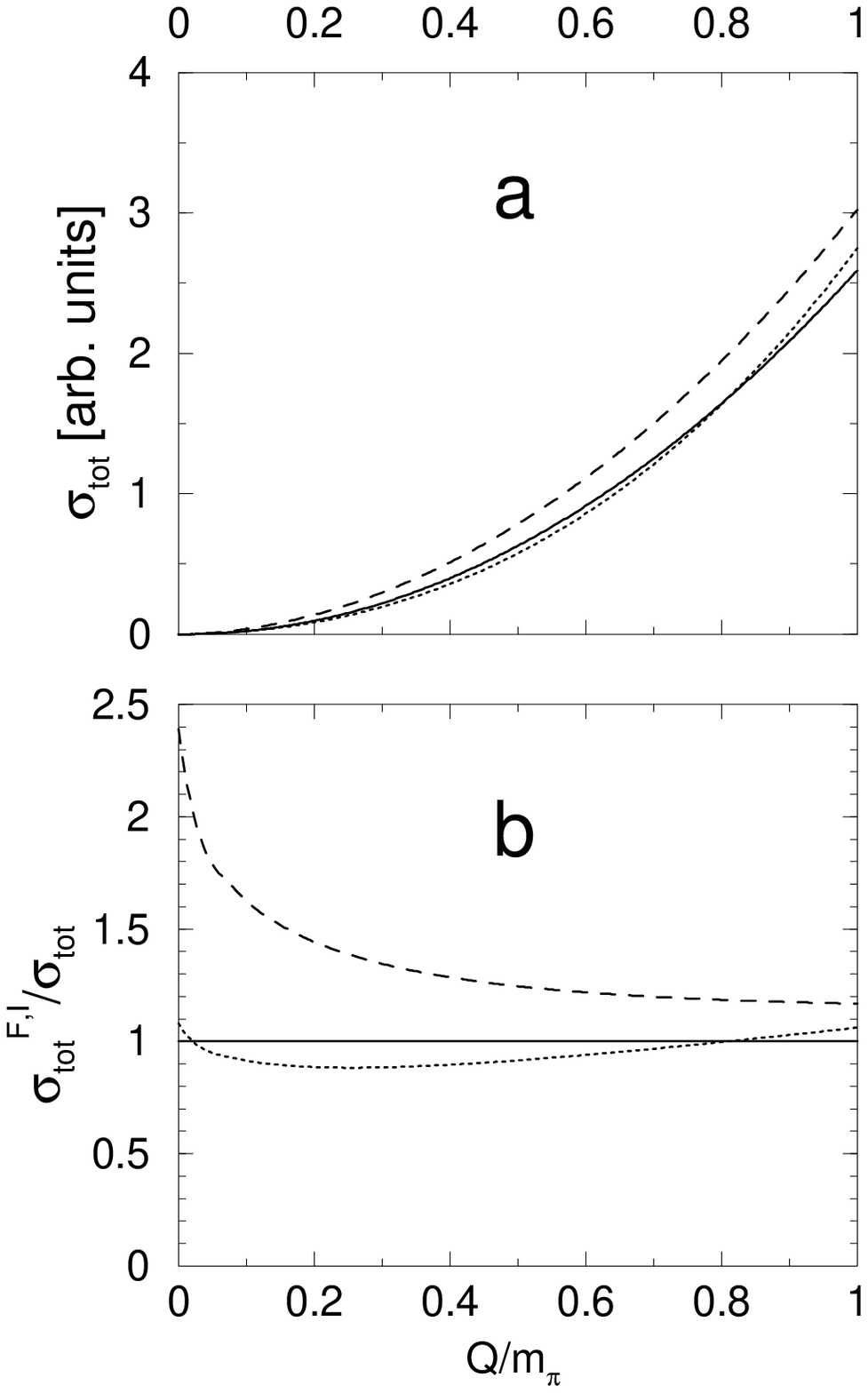,height=10.5cm,width=7.25cm}}

\newpage

{\bf Fig.~5}

\centerline{\psfig{file=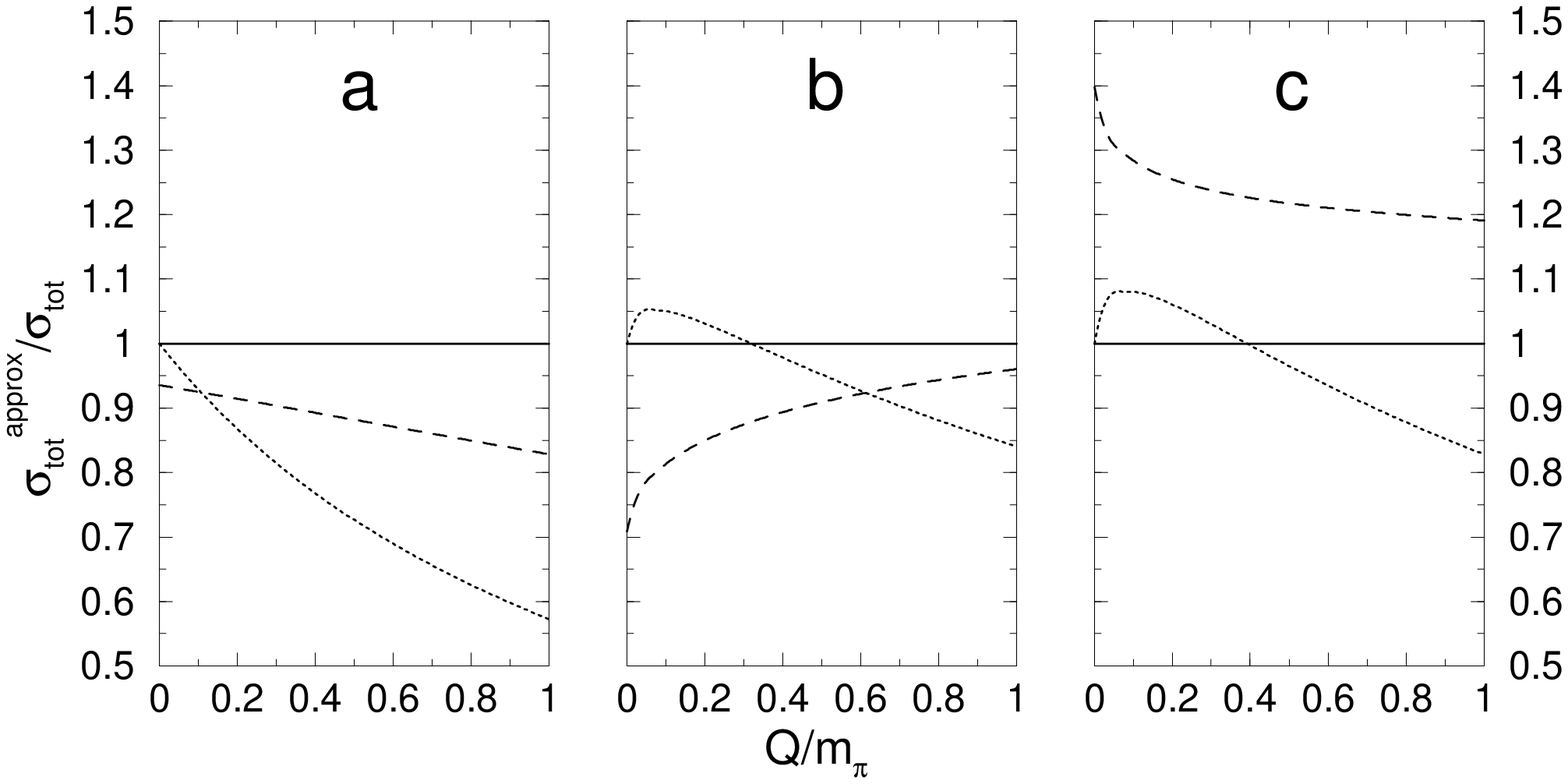,height=7.00cm,width=14.5cm}}

\vspace{20mm}

{\bf Fig.~6}

\centerline{\psfig{file=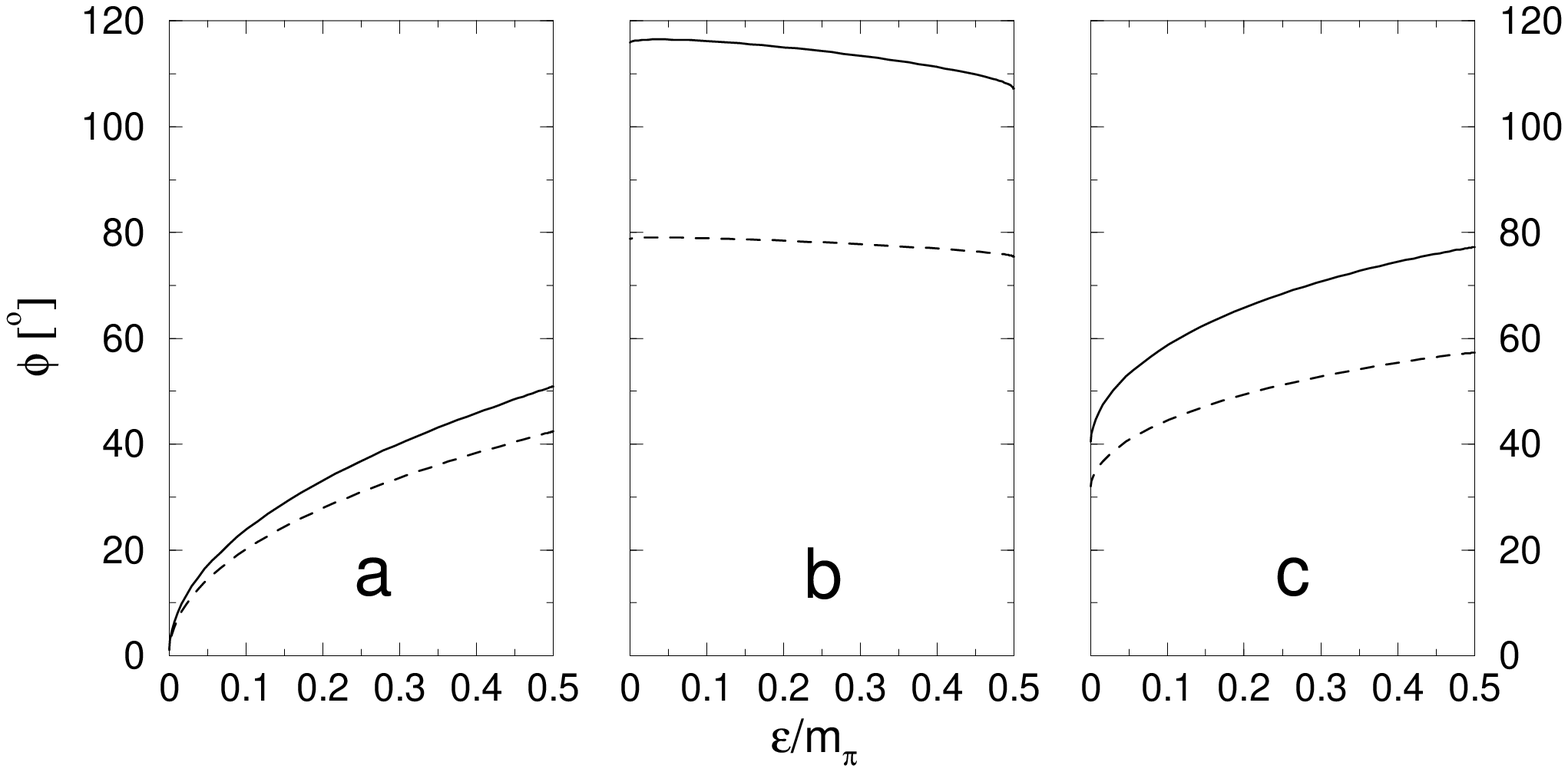,height=7.00cm,width=14.5cm}}

\end{document}